\input harvmac.tex
\noblackbox
\input epsf.sty


\font\cmss=cmss10
\font\cmsss=cmss10 at 7pt

\def\inbar{\vrule height1.5ex width.4pt depth0pt}

\def\IN{\relax{\rm I\kern-.18em N}}
\def\IB{\relax\hbox{$\inbar\kern-.3em{\rm B}$}}
\def\IC{\relax\hbox{$\inbar\kern-.3em{\rm C}$}}
\def\IQ{\relax\hbox{$\inbar\kern-.3em{\rm Q}$}}
\def\ID{\relax\hbox{$\inbar\kern-.3em{\rm D}$}}
\def\IE{\relax\hbox{$\inbar\kern-.3em{\rm E}$}}
\def\IF{\relax\hbox{$\inbar\kern-.3em{\rm F}$}}
\def\IG{\relax\hbox{$\inbar\kern-.3em{\rm G}$}}
\def\IGa{\relax\hbox{${\rm I}\kern-.18em\Gamma$}}
\def\IH{\relax{\rm I\kern-.18em H}}
\def\IK{\relax{\rm I\kern-.18em K}}
\def\IL{\relax{\rm I\kern-.18em L}}
\def\IP{\relax{\rm I\kern-.18em P}}
\def\IR{\relax{\rm I\kern-.18em R}}
\def\Z{\relax\ifmmode\mathchoice{\hbox{\cmss Z\kern-.4em Z}}{\hbox{\cmss Z\kern-.4em Z}} {\lower.9pt\hbox{\cmsss Z\kern-.4em Z}}{\lower1.2pt\hbox{\cmsss Z\kern-.4em Z}}\else{\cmss Z\kern-.4em Z}\fi}

\def\II{\relax{\rm I\kern-.18em I}}
\def\one{\relax{\rm 1\kern-.25em I}}

\def\CLL{\relax{\CL\kern-.74em \CL}}

\def\CC{{\cal C}}

\def\CN{{\cal N}}

\def\CW{{\cal W}}

\def\CN#1{{\cal N}=#1}


\lref\ABSV{
  M.~Aganagic, C.~Beem, J.~Seo and C.~Vafa,
  ``Geometrically Induced Metastability and Holography,''
  [hep-th/0610249].
}

\lref\sinha{
  S.~Sinha and C.~Vafa,
  ``SO and Sp Chern-Simons at large N,''
  arXiv:hep-th/0012136.
}

\lref\AVP{
  M.~Aganagic and C.~Vafa,
  ``Perturbative derivation of mirror symmetry,''
  arXiv:hep-th/0209138.
}

\lref\HSV{
  J.~J.~Heckman, J.~Seo and C.~Vafa,
  ``Phase Structure of a Brane/Anti-Brane System at Large N,''
  JHEP {\bf 0707}, 073 (2007)
  [arXiv:hep-th/0702077].
}

\lref\ABF{
  M.~Aganagic, C.~Beem and B.~Freivogel,
  ``Geometric Metastability, Quivers and Holography,''
  arXiv:0607.0596 [hep-th].
}

\lref\Kachru{
  O.~Aharony and S.~Kachru,
  ``Stringy Instantons and Cascading Quivers,''
  arXiv:0707.3126 [hep-th].
}

\lref\AKS{
O. Aharony, S. Kachru and E. Silverstein,
``Simple Stringy Dynamical SUSY Breaking,''
arXiv:0708.0493 [hep-th].}

\lref\Vafa{
C. Vafa, ``Superstrings and topological strings at large N,''
arXiv:hep-th/0008142.}

\lref\ABFK{
R. Argurio, M. Bertolini, S. Franco and S. Kachru,
``Metastable vacua and D-branes at the conifold,''
arXiv:hep-th/0703236.}

\lref\DFS{
M. Dine, J. Feng and E. Silverstein,
``Retrofitting O'Raifeartaigh models with dynamical scales,"
arXiv:hep-th/0608159.}

\lref\Kumar{J. Kumar, ``Dynamical SUSY breaking in intersecting brane models,''
arXiv:0708.4116.}

\lref\reviews{
For excellent reviews of dynamical SUSY breaking in 4d quantum field theory, see:
I. Affleck, M. Dine and N. Seiberg, ``Dynamical supersymmetry breaking in four-dimensions
and its phenomenological implications," Nucl. Phys. {\bf B256} (1985) 557;
E. Poppitz and S.P. Trivedi, ``Dynamical supersymmetry breaking," Ann. Rev. Nucl. Part. Sci.
{\bf 48} (1998) 307, arXiv:hep-th/9803107;
Y. Shadmi and Y. Shirman, ``Dynamical supersymmetry breaking," Rev. Mod. Phys. {\bf 72} (2000) 25,
arXiv:hep-th/9907225;
K. Intriligator and N. Seiberg, ``Lectures on supersymmetry breaking," arXiv:hep-th/0702069.}

\lref\GVW{
S. Gukov, C. Vafa and E. Witten, ``CFTs from Calabi-Yau Fourfolds,''
arXiv:hep-th/9906070.}

\lref\Mina{
  M.~Aganagic and C.~Vafa,
  ``Mirror Symmetry, D-branes and Counting Holomorphic Discs,''
  [hep-th/0012041].
}
\lref\AAHV{
  B.~Acharya, M.~Aganagic, K.~Hori and C.~Vafa,
  ``Orientifolds, mirror symmetry and superpotentials,''
  arXiv:hep-th/0202208.
}

\lref\Witten{
  E.~Witten,
  ``Branes and the Dynamics of QCD,''
  [hep-th/9706109].
}
\lref\AMV{
  M.~Atiyah, J.~M.~Maldacena and C.~Vafa,
  ``An M-theory flop as a large N duality,''
  J.\ Math.\ Phys.\  {\bf 42}, 3209 (2001)
  [arXiv:hep-th/0011256].
}

\lref\Blumenhagen{N. Akerblom, R. Blumenhagen, D. Lust, E. Plauschinn and M. Schmidt-Sommerfeld,
``Non-perturbative SQCD superpotentials from String Instantons,'' JHEP {\bf 0704} (2007) 076,
arXiv:hep-th/0612132.}

\lref\sinst{R. Blumenhagen, M. Cvetic and T. Weigand, ``Spacetime instanton corrections in 4D string vacua:
The seesaw mechanism for D-brane models," Nucl. Phys. {\bf B771} (2007) 113, arXiv:hep-th/0609191;
M. Haack, D. Krefl, D. Lust, A. Van Proeyen and M. Zagermann, ``Gaugino condensates and D-terms
for D7-branes," JHEP {\bf 0701} (2007) 078, arXiv:hep-th/0609211;
L. Ibanez and A. Uranga, ``Neutrino majorana masses from string theory instanton effects,"
JHEP {\bf 0703} (2007) 052, arXiv:hep-th/0609213;
B. Florea, S. Kachru, J. McGreevy and N. Saulina, ``Stringy instantons and quiver gauge theories,"
JHEP {\bf 0705} (2007) 024, arXiv:hep-th/0610003.}

\lref\zeromodes{M. Bianchi and E. Kiritsis,
``Non-perturbative and flux superpotentials for Type I
strings on the $Z_3$ orbifold, arXiv:hep-th/0702015;
R. Argurio, M. Bertolini, G. Ferretti, A. Lerda
and C. Petersson, ``Stringy instantons at orbifold
singularities," arXiv:0704.0262 [hep-th];
M. Bianchi, F. Fucio and J.F. Morales, ``D-brane
instantons on the $T^6/Z_3$ orientifold,''
arXiv:0704.0784 [hep-th];
L. Ibanez, A. Schellekens and A. Uranga,
``Instanton induced neutrino majorana masses
in CFT orientifolds with MSSM-like spectra,''
arXiv:0704.1079 [hep-th];
R. Blumenhagen and M. Cvetic, ``Lifting D-instanton
zero modes by recombination and background fluxes,''
arXiv:0708.0403 [hep-th].}

\lref\AV{
  M.~Aganagic and C.~Vafa,
  ``Mirror symmetry and a G(2) flop,''
  JHEP {\bf 0305}, 061 (2003)
  [arXiv:hep-th/0105225].
}

\lref\AW{
  M.~Atiyah and E.~Witten,
  ``M-theory dynamics on a manifold of G(2) holonomy,''
  Adv.\ Theor.\ Math.\ Phys.\  {\bf 6}, 1 (2003)
  [arXiv:hep-th/0107177].
}

\lref\DV{
  R.~Dijkgraaf and C.~Vafa,
  ``A perturbative window into non-perturbative physics,''
  arXiv:hep-th/0208048.
}

\lref\CKV{
  F.~Cachazo, S.~Katz and C.~Vafa,
  ``Geometric transitions and N = 1 quiver theories,''
  arXiv:hep-th/0108120.
}

\lref\hori{
  I.~Brunner and K.~Hori,
  ``Orientifolds and mirror symmetry,''
  JHEP {\bf 0411}, 005 (2004)
  [arXiv:hep-th/0303135].
}

\lref\ISS{
K. Intriligator, N. Seiberg and D. Shih, ``Dynamical SUSY breaking in meta-stable
vacua," arXiv:hep-th/0602239.}

\lref\HV{
  J.~J.~Heckman and C.~Vafa,
  ``Geometrically Induced Phase Transitions at Large N,''
  arXiv:0707.4011 [hep-th].
}

\lref\IKV{
  K.~Intriligator, P.~Kraus, A.~V.~Ryzhov, M.~Shigemori and C.~Vafa,
  ``On low rank classical groups in string theory, gauge theory and 
  matrix models,''
  Nucl.\ Phys.\  B {\bf 682}, 45 (2004)
  [arXiv:hep-th/0311181].
}

\lref\lazaroiu{
  K.~Landsteiner, C.~I.~Lazaroiu and R.~Tatar,
  ``(Anti)symmetric matter and superpotentials from IIB orientifolds,''
  JHEP {\bf 0311}, 044 (2003)
  [arXiv:hep-th/0306236].
}

\lref\Do{
  M.~R.~Douglas, J.~Shelton and G.~Torroba,
  ``Warping and supersymmetry breaking,''
  arXiv:0704.4001 [hep-th].
}

\lref\oz{
  H.~Ita, H.~Nieder and Y.~Oz,
  ``Perturbative computation of glueball superpotentials for SO(N) and
 USp(N),''
  JHEP {\bf 0301}, 018 (2003)
  [arXiv:hep-th/0211261].
}

\lref\phases{
  E.~Witten,
  ``Phases of N = 2 theories in two dimensions,''
  Nucl.\ Phys.\  B {\bf 403}, 159 (1993)
  [arXiv:hep-th/9301042].
}

\lref\KS{
  I.~R.~Klebanov and M.~J.~Strassler,
  ``Supergravity and a confining gauge theory: Duality cascades and
  chiSB-resolution of naked singularities,''
  JHEP {\bf 0008}, 052 (2000)
  [arXiv:hep-th/0007191].
}

\lref\MN{
  J.~M.~Maldacena and C.~Nunez,
  ``Towards the large N limit of pure N = 1 super Yang Mills,''
  Phys.\ Rev.\ Lett.\  {\bf 86}, 588 (2001)
  [arXiv:hep-th/0008001].
}

\lref\ADS{I. Affleck, M. Dine and N. Seiberg, ``Dynamical supersymmetry breaking in supersymmetric QCD,"
Nucl. Phys. {\bf B241} (1984) 493.}

\lref\Giveon{A. Giveon and D. Kutasov, ``Gauge symmetry and supersymmetry breaking from intersecting
branes," Nucl. Phys. {\bf B778} (2007) 129 [arXiv:hep-th/0703135].}

\lref\conifold{
H. Ooguri and C. Vafa, ``Summing up D-instantons,''
arXiv:hep-th/9608079;
N. Seiberg and S. Shenker, ``Hypermultiplet moduli space and string
compactifications to three-dimensions,'' arXiv:hep-th/9608086.}

\lref\Nima{N.~Arkani-Hamed, M.~Dine and S.~P.~Martin, ``Dynamical supersymmetry breaking in models with a Green-Schwarz mechanism,''
Phys.\ Lett.\  B {\bf 431}, 329 (1998) [arXiv:hep-ph/9803432].}


\Title{\vbox{\baselineskip12pt\hbox{SLAC-PUB-12821, SU-ITP-07/15}
\hbox{} }}
 {\vbox{ {\centerline{Geometric Transitions and Dynamical SUSY Breaking} }}}

\centerline{Mina Aganagic,$^{a}$ Christopher Beem,$^{a}$ and Shamit Kachru$^{b}$}
\bigskip
{\it \centerline{$^{a}$Department of Physics, University of California, Berkeley, CA 94720}
{\it \centerline{$^{b}$Department of Physics and SLAC, Stanford University, Stanford CA 94305}}
}
\smallskip

\bigskip
\noindent

We show that the physics of D-brane theories that exhibit dynamical SUSY breaking due to stringy instanton effects is well captured by geometric transitions, which recast the non-perturbative superpotential as a classical flux superpotential.  This allows for simple engineering of Fayet, Polonyi, O'Raifeartaigh, and other canonical models of supersymmetry breaking in which an exponentially small scale of breaking can be understood either as coming from stringy instantons or as arising from the classical dynamics of fluxes.

\Date{September 2007}

\newsec{Introduction}

It is of significant interest to find simple examples of dynamical supersymmetry breaking in string theory.  One class of examples,
where stringy D-instanton effects play a starring role, was described in \AKS.
These models exhibit ``retrofitting" of the classic SUSY breaking theories (Fayet, Polonyi and O'Raifeartaigh) \DFS, without incorporating
any nontrivial gauge dynamics.  Instead, stringy instantons \sinst\ automatically implement the exponentially small scale of SUSY breaking
in theories with only Abelian gauge fields.
A related idea using disc instantons instead of D-instantons appears in \Kumar.
These models are simpler in many ways than their existing field theory analogues \reviews.

In this paper, we show that these results (and many generalizations) admit a clear and computationally powerful understanding
using geometric transition techniques \Vafa (see also \refs{\KS,\MN}).
Such techniques are well known to translate quantum computations of superpotential interactions in non-trivial gauge theories to classical geometric computations of flux-induced superpotentials \GVW.
They are most powerful when the theories in question exhibit a mass gap.  While the classic models we study {\it do} manifest light degrees of freedom (and hence do not admit a complete description in terms of geometry and fluxes), we find that a mixed description involving small
numbers of D-branes in a flux background -- which arises after a geometric transition from a system of branes at a singularity --
nicely captures the relevant physics of supersymmetry breaking\foot{For an application of geometric transitions to the study of supersymmetry breaking in the context of brane/anti-brane systems see
\refs{\ABSV\HSV\HV\ABF-\Do}.}. In the original theory without flux, the SUSY breaking effects are generated by D-instantons either in $U(1)$ gauge factors or on unoccupied, but orientifolded, nodes of the quiver
gauge theory (analogous to those studied in
\refs{\AKS,\Kachru,\ABFK}).  Both are in some sense ``stringy''
effects.  Simple generalizations involve more familiar transitions on
nodes with large $N$ gauge groups.

The geometric transition techniques we apply have two advantages over the description using stringy instantons in a background without fluxes.  First, they allow for a classical computation of the relevant superpotential instead of requiring a nontrivial instanton
calculation.  Second, they incorporate higher order corrections (due
to multi-instanton effects in the original description) which had not
been previously calculated.

The organization of this paper is as follows.  In section 2, we remind
the reader of the relevant background about geometric transitions.  In
section 3, we discuss the geometries we will use to formulate our DSB
theories.  In sections 4-6, we give elementary examples that yield
Fayet, Polonyi, and O'Raifeartaigh models that break SUSY at
exponentially low scales.  In section 7, we present a single geometry
that unifies the three models, reducing to them in various limits.  In 
section 8, we provide a more general, exact analysis of the existence
of these kinds of susy-breaking effects.  In section 9, we give a few 
other examples of simple DSB theories
(related to recent or well known literature in the area).  Finally, in
section 10, we extend the technology to orientifold models, in
particular recovering models which are closely related to the
specific examples of \AKS.

\newsec{Background: Geometric Transitions}

Computing non-perturbative corrections in string theory, even to
holomorphic quantities such as a superpotential, is in general very
difficult. A surprising recent development \refs{\Vafa,\DV}
is that in some cases -- namely for massive theories --
these non-perturbative effects can be determined by perturbative means
in a dual language\foot{For a two-dimensional example see \AVP .}.

Consider, for example, $N$ D5 branes in type IIB string theory
wrapping an isolated, rigid $\IP^1$ in a local Calabi-Yau manifold.
In the presence of D5 branes, D1 brane instantons wrapping the $\IP^1$
generate a superpotential for the K\"ahler moduli\foot{This is a slight misnomer, since $t$ is a parameter, and not a dynamical field for a non-compact Calabi-Yau.}.
The instanton
effects are proportional to
$$
\exp\left( -{t\over N g_s}\right)
$$
where
$
t = \int_{S^2} (B^{NS} + i g_s B^{RR}).
$
For general $N$, these D1 brane instantons are gauge theory instantons.  
More precisely, they are the fractional $U(N)$ instantons of the low energy ${\cal N}=1$ $U(N)$ gauge theory on the D5 brane. However, on the basis of zero mode counting, one expects that stringy instanton effects are present even for a single D5 brane.

In the absence of D5 branes, the theory has $\CN2$ supersymmetry, and
the K\"ahler moduli space is unlifted. In that case, the local Calabi-Yau with a rigid $\IP^1$ is known to have another phase where the $S^2$ has shrunk to zero size and has been replaced by a finite $S^3$. The two branches meet at $t=0$, where there is a singularity at which the D3 branes wrapping the $S^3$ become massless.

What happens to this phase transition in the presence of D5 branes? Classically, we can still connect the $S^2$ to the $S^3$ side by a geometric transition. The only difference is that to account for the D5 brane charge, we need there to be $N$ units of RR flux through the 
$S^3$,
$$
\int_{S^3} H^{RR} = N.
$$
Quantum mechanically the effect is more dramatic.  In the presence of
D5 branes there is no sharp phase transition at all between the $S^2$ and the $S^3$ sides; the interpolation between them is completely
$smooth$.  As a consequence, the two sides of the transition provide
$dual$ descriptions of the same physics. Since the theory is massive now, the interpolation occurs by varying the coupling constants of the theory. The fact that the singularity where the $S^3$ shrinks to zero size is eliminated is consistent with the fact that D3 branes wrapping an $S^3$ with RR flux through it are infinitely massive. The most direct proof of the absence of a phase transition is in the context of M-theory on a $G_2$ holonomy manifold \refs{\AMV,\AV,\AW}.
This is related to the present transition by mirror symmetry
and an M-theory lift. In M-theory, the transition is analogous to a
perturbative flop transition of type IIA string theory at the
conifold, except that in M-theory the classical geometry gets
corrected by M2 brane instantons instead of worldsheet instantons
\AMV. The argument that the two sides are connected smoothly is
analogous to Witten's argument for the absence of a sharp phase
transition in IIA \phases. In both cases, the presence of instantons
is crucial for the singularities in the interior of the classical moduli space to be eliminated.

The fact that the two sides of the transition are connected smoothly implies that the superpotentials have to be the same. The instanton-generated superpotential has a dual description on the $S^3$ side as a
$perturbative$ superpotential generated by fluxes. The flux superpotential
$$
{\cal W} = \int H \wedge \Omega
$$
is perturbative, given by
\eqn\sup{
{\cal W} = {t\over g_s} \; S + N \; \del_{S} {\cal F}_0
}
where ${\cal F}_0(S)$ is the prepotential of the Calabi-Yau,
and
$$
S = \int_{S^3} \Omega.
$$
The first term in \sup\ comes from the running of the gauge coupling
$t/g_s$ which implies that there is an $H^{NS}$ flux turned on on the
Calabi-Yau through a 3-chain on the $S^2$ side.  This three-chain
becomes the non-compact 3-cycle dual to the $S^3$ after the transition.
Near the conifold
$$
{\del_S} {\cal F}_0 = S\left(\log\left({S\over\Delta^3}\right) -1\right) + \ldots
$$
where the omitted terms are a model dependent power series in $S$, and $\Delta$ is a high scale at which $t$ is defined.
Integrating out $S$ in favor of $t$, the superpotential ${\cal W}$ becomes
$$
\CW_{inst}= - \Delta^3 exp(-{t\over N g_s}) + \ldots
$$
up to two and higher order instanton terms that depend on the
power series in ${\cal F}_0(S).$
The duality should persist even in the presence of other branes and fluxes, as long as the $S^2$ that the branes wrap remains isolated,
and the geometry near the branes is unaffected.
As we'll discuss in section 10, this can also be extended to D5 branes wrapping $\IP^1$'s
in Calabi-Yau orientifolds.

\newsec{The Theories}

To construct the models in question, we will consider type IIB on non-compact Calabi-Yau 3-folds which are $A_r$ ADE type ALE spaces fibered over the complex plane $~\IC[x]$.
These are described as hypersurfaces in $~\IC^4$ as follows
\eqn\ALE{
uv=\prod_{i=1}^{r+1}(z-z_i(x)).
}
This geometry is singular at points where $u,v=0$ and $z_i(x)=z_j(x)=z$.  At these points, there are vanishing size $\IP^1$'s which can be blown up by deforming the K\"ahler parameters of the Calabi-Yau.
There are $r$ 2-cycle classes, which we will denote
$$
S^2_i.
$$
These correspond to the blow-ups of the singularities at $z_i=z_{i+1}$, $i=1,\ldots r$.
It is upon these $\IP^1$'s that we wrap D5 branes to engineer our gauge theories.

The theory on the branes can be thought of as an ${\cal N}=2$ theory,
corresponding to D5 branes wrapping 2-cycles of the ALE space, which is then
deformed to an ${\cal N}=1$ theory by superpotentials for the adjoints.
For the branes on $S^2_i$ this superpotential is denoted $\CW_i(\Phi_i)$.
The adjoint $\Phi_i$ describes
the positions of the branes in the $x$-direction, and the superpotential arises because the ALE space is fibered nontrivially over the $x$ plane.
The superpotential can be computed by integrating \refs{\Witten, \Mina}
$$
\CW =\int_{\CC} \Omega
$$
over a 3-chain with one boundary as the wrapped $S^2$.
In this particular geometry, it takes an extra simple form
(the details of the computation appear in appendix A)
\eqn\super{
\CW_i(x)=\int (z_i(x)-z_{i+1}(x))dx.
}

In addition to the adjoints, for each intersecting pair of two-cycles $S_i^2$, $S^2_{i+1}$
there is a bifundamental hypermultiplet at the intersection, consisting of chiral multiplets $Q_{i, i+1}$
and ${Q}_{i+1,1}$, with a superpotential interaction inherited from the ${\cal N}=2$ theory
$$
{\rm Tr}(Q_{i, i+1}\Phi_{i+1} Q_{i+1,i}-Q_{i, i+1}Q_{i+1,i}\Phi_i).
$$

Classically, the vacua of the theory correspond to
the different ways of distributing branes on the minimal
$\IP^1$'s in the geometry \CKV . When one of the nodes is massive, the instantons corresponding to D1 branes wrapping the $S^2$ can be summed up in the dual geometry after a geometric transition.  As explained in \AKS, and as we'll see in
the next section, this can trigger supersymmetry breaking in the rest of the system.

As an aside, we note that the systems we are studying are a slight generalization of those described in \refs{\Kachru,\AKS}.
Those geometries are related to the family of geometries studied here, but correspond to particular points in the parameter space where the adjoint masses have been taken to be large and the branes and/or O-planes have been taken to coincide in the $x$-plane.
In addition, we allow the possibility of $U(1)$ (or in some cases higher rank) gauge groups on the transitioning node, whereas in \refs{\Kachru,\AKS} the instanton effects were associated with nodes that were only occupied by O-planes.
Nevertheless, we will find the same qualitative physics as in \AKS\ in this broader class of theories.

\newsec{The Fayet Geometry}

We now turn to the specific geometry which will engineer the Fayet model at low energies.  This will be an $A_3$ geometry, and \ALE\ can be written explicitly as
\eqn\athree{
uv=(z-mx)(z+mx)(z-mx)(z+m(x-2a)).
}

After blowing up, we wrap $M$ branes each on $S^2_1$ at $z_1(x)=z_2(x)$,
on $S^2_2$ at $z_2(x)=z_3(x)$ and one brane on $S^2_3$ at $z_3(x)=z_4(x)$. The
tree-level superpotential \super\ is now given by
\eqn\tree{\eqalign{ \CW= \sum_{i=1}^3\CW_i(\Phi_i) +
{\rm Tr}(Q_{12}\Phi_2 Q_{21} - Q_{21}\Phi_1 Q_{12}) +
{\rm Tr}(Q_{23}\Phi_3 Q_{32}- Q_{32}\Phi_2 Q_{23})}} where
$$
\CW_1(\Phi_1) = m \Phi_1^2, \qquad \CW_2(\Phi_2)=-m \Phi_2^2,\qquad
\CW_3(\Phi_3)=m(\Phi_3-a)^2.
$$
%
\bigskip
\centerline{\epsfxsize 3.3truein\epsfbox{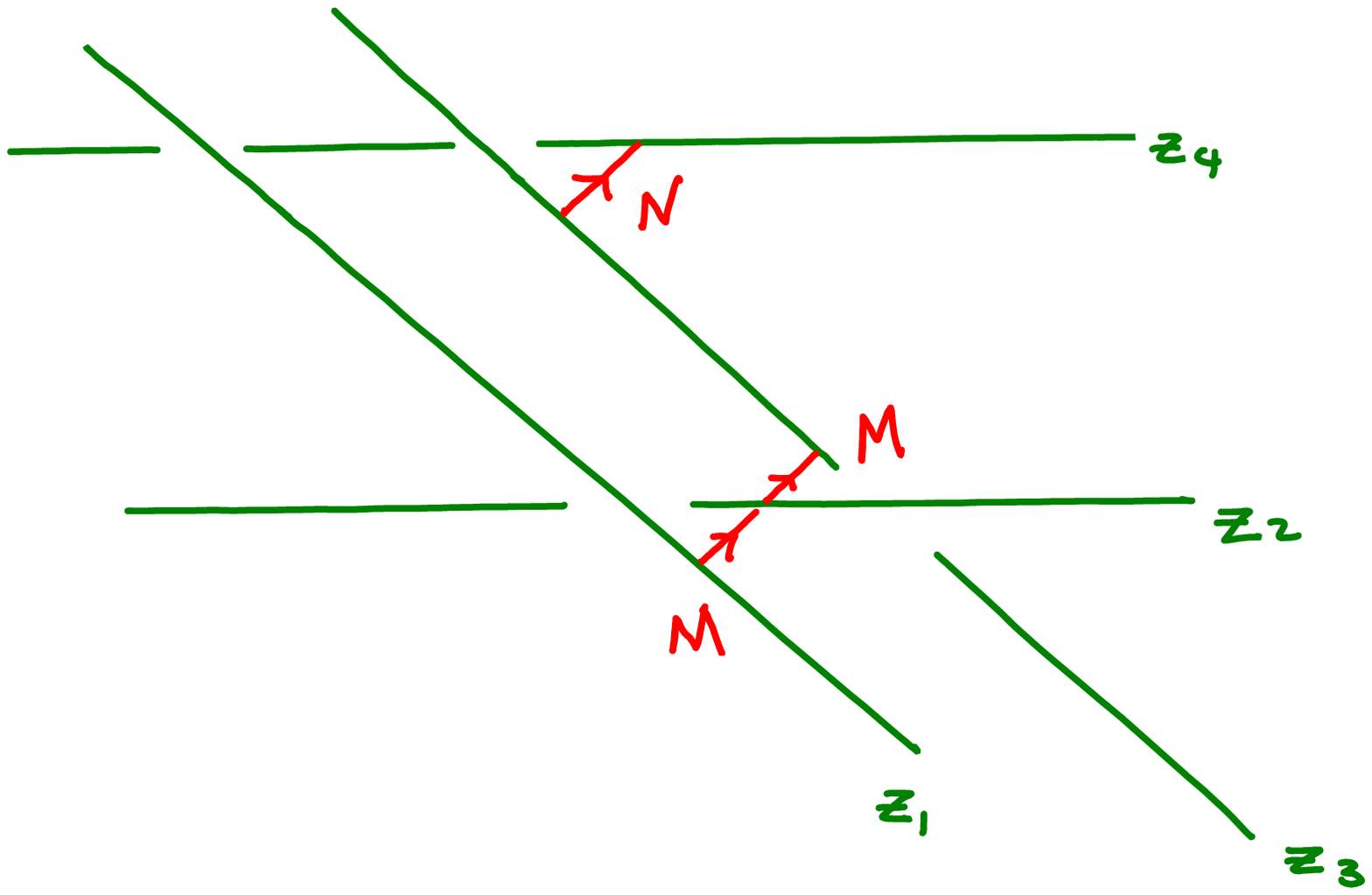}}
\noindent{\ninepoint
\baselineskip=2pt {\bf Fig. 1.} {The $A_3$ geometry used for retrofitting the Fayet model, {\it before} the geometric transition. The red lines represent the $\IP^1$'s, wrapped by D5 branes. The third node does not intersect the other two and is massive. The geometry after the transition sums up the corresponding instantons. For $N=1$ branes on $S^2_3$, the instantons are stringy. For $N>1$, these are fractional instantons associated with gaugino condensation in the pure $U(N)$ ${\cal N}=1$ gauge theory on that node. }}
\bigskip
The branes on nodes one and two intersect, since both of the
corresponding $\IP^1$'s are at $x=0$.  However, the third node, and the single brane on it, is isolated at $x=a$, and the theory living on it is massive.  Correspondingly, the the instantons effects due to D-instantons wrapping the third node can be summed up in a dual geometry where we trade $S^2_3$ for a three-cycle $S^3$
with one unit of flux through it
$$
\int_{S^3} H^{RR} = 1~.
$$

The geometry after the transition is described by the deformed equation
\eqn\dual{
uv=(z-mx)(z+mx)((z-mx)(z+m(x-2 a))-s)
}
where the size of the $S^3$
$$
\int_{S^3} \Omega = S
$$
is given by $S= s/m$. It is fixed to be exponentially small by the flux
superpotential, as we shall see shortly.  The third brane is gone now, and so are the
fields $Q_{23}$, $Q_{32}$ and $\Phi_3$.
The effective superpotential can now be written to leading order in $S$ as
$$
{\CW}_{eff}= \CW_1(\Phi_1)+{\tilde W}_2(\Phi_2,S) +
{\rm Tr}(Q_{12}\Phi_2Q_{21}-Q_{21}\Phi_1Q_{12})+ \CW_{flux}(S).
$$
In this geometry, the exact flux superpotential is
$$
{\CW}_{flux}=
{t\over g_s} S+ S\left(\log{S\over\Delta^3}-1\right)
$$
without any polynomial corrections in $S$.  It is crucial here that the
superpotential for $\Phi_2$ has changed due to the change in the geometry to
$
{\tilde \CW}_2(\Phi_2)
$,
where
$$
{\tilde \CW}_2(x)=\int (z_2(x)-\tilde z_3(x))dx,
$$
while the superpotential for $\Phi_1$ is unaffected.
We have defined
$$
(z-\tilde z_{3}(x))(z-\tilde z_{4}(x))=(z-z_3(x))(z-z_4(x))-s
$$
with $\tilde z_3(x)$ being the branch which asymptotically looks like $z_3(x)$ at
large values of $x$. In other words,
$$
{\tilde \CW}_2(x)=\int^{x}_{\Delta} (-m(x'+a)-\sqrt{m^2(x'-{a})^2+s})dx'.
$$
This superpotential sums up the instanton effects due to Euclidean branes wrapping node
three.

Before the transition, the vacuum was at ${\Phi_2}=0$.
At the end of the day, we expect it to be perturbed by exponentially small terms $\sim S$, so the relevant part of the superpotential is
\eqn\modsuper{
{\tilde W}_2(\Phi_2)=-m{\rm Tr}\Phi_2^2-
{1\over2}\,S\,{\rm Tr} \log{a-\Phi_2\over\Delta}+\ldots.
}
where we've omitted terms of order $S^2$ and higher and dropped an
irrelevant constant.  We comment on the form of these corrections
in appendix B.

The theories on nodes one and two are asymptotically free.
If the fields $S$ and $\Phi_{1,2}$ have very large masses, we can
integrate them out and keep only the light degrees of freedom.  Keeping only the leading instanton corrections, the relevant F-terms are
\eqn\fterms{\eqalign{
&F_{\Phi_1}=2m\Phi_1- Q_{12} Q_{21}\cr
&F_{\Phi_2}=-2m\Phi_2+ Q_{21} Q_{12}+{S\over2(a-\Phi_2)}\cr
&F_{S}=t/g_s+\log{S/\Delta^3}-{1\over2}{\rm Tr}\log(a-\Phi_2)/\Delta
}}
Setting these to zero, we obtain
$$S_*=\Delta^3
\exp({-{{\tilde t}\over g_s}})+\ldots
$$
where
$$
\tilde t = t - {1\over 2} M g_s \log(a/\Delta)
$$
and
\eqn\ftermsols{\eqalign{
&\qquad\qquad\qquad\Phi_{1,*} =-{1\over2m}Q_{12}Q_{21}\cr
&\qquad\qquad\qquad\Phi_{2,*} ={1\over2m}Q_{21}Q_{12}+
{1\over 4 m a} S_{*} +\ldots.
}}
The omitted terms are higher order in $Q_{21}Q_{12}/ma$ and
$\exp(-{{\tilde t}\over g_s}).$
The low energy, effective superpotential is
$$
\CW_{eff}={1\over m}{\rm Tr}(Q_{12}Q_{21}Q_{12}Q_{21})~
 -\,{S_*\over4 ma}{\rm Tr}Q_{12} Q_{21} + \ldots $$
where we have neglected corrections to the quartic coupling, and the
higher order couplings of $Q's$, all of which are exponentially
suppressed.
As shown in \AKS, in the presence of a generic FI
term for the off-diagonal $U(1)$ under which $Q_{12}$ and
$Q_{21}$ are charged,
$$
D = Q_{12} Q_{12}^\dagger -Q_{21}^\dagger Q_{21} - r,
$$
the exponentially small
mass for $Q$ will trigger F-term supersymmetry breaking with an
exponentially low scale; we can put $Q_{12,*} = \sqrt r $, and then
$$
F_{Q_{21}} \sim {\sqrt r \over4 ma} S_{*}~.
$$
Geometrically, turning the FI term corresponds to choosing the central
charges of the branes on the two nodes to be miss-aligned.  Combined
with the fact that the nodes one and two have become massive with an
exponentially low mass, this provides an extremely simple mechanism of
breaking supersymmetry at a low scale.  The non-supersymmetric vacuum
we found classically is reliable, as long as the scale of
supersymmetry breaking is far above the strong coupling scales of the
$U(M)\times U(M)$ gauge theory.  Had we taken $N$ branes on the
massive node instead of one, the story would have been the same, apart
from the fact that the flux increases, and correspondingly the vacuum
value of $S$ changes to $S_*\sim\Delta^3
\exp({-{{\tilde t}\over N g_s}})$. In this case however, the instantons that trigger supersymmetry breaking are the fractional $U(N)$ instantons.

\newsec{The Polonyi Model}

In this section we construct the Polonyi model with an exponentially
small linear superpotential term for a chiral superfield $\Phi$.  This
will turn out to be somewhat more subtle, and the existence of the (meta)stable vacuum will depend sensitively on the K\"ahler potential.
We describe specific cases where we know the relevant K\"ahler potential does yield a stable vacuum in section 7.

Consider an $A_2$ geometry given by
\eqn\Polgeom{
uv=(z-mx)(z-mx)(z+m(x-2a))
}
which has one D5-brane wrapped on the $S^2_1$ blowing up
$z_1(x)=z_2(x)$, and one D5-brane wrapped on the $S^2_2$ blowing up $z_2(x)=z_3(x)$.  This system has a tree-level superpotential
\eqn\superpol{
\CW=
 \CW_{1}(\Phi_1)  +\CW_{2}(\Phi_2)
+Q_{12}\Phi_2 Q_{21}- Q_{21}\Phi_1Q_{12}.
}

where
$$
\CW_{1}(\Phi_1)=0,  \qquad \CW_{2}(\Phi_2) = m(\Phi_2 - a)^2
$$
This theory has a classical moduli space of vacua
parameterized by the expectation value of $\Phi_1$ and where $Q_{12,*}=0=Q_{21,*}$,
and ${\Phi}_{2,*}=a$.

At a generic point in the moduli space, away from
$\Phi_1 =a$, the theory on the branes wrapping $S^2_2$ is massive.
Then, the instanton effects associated with D1 branes wrapping this node
can be summed up by a geometric transition, that replaces $S^2_2$ by an $S^3$ with one unit of flux through it. This deforms the Calabi-Yau geometry to
$$
uv=(z-mx)((z-mx)(z+m(x-2a))-s).
$$
which has an $S^3$ of size
$$
\int_{S^3} \Omega = S
$$
where $S=s/m$. With this deformation, the superpotential for node 1 is altered as well:
$$
{\tilde \CW}_1(x)=\int( -m(a-x)+\sqrt{m^2(a-{x})^2+s})dx.
$$

The effective superpotential after the transition is simply
$$
\CW_{eff} = {\tilde W}_1(\Phi_1,S)+
\CW_{flux}(S)
$$
where the flux superpotential has the simple form:
$$
\CW_{flux}(S) = {t\over g_s}\, S + S(\log{S/\Delta^3}-1)
$$
Note that there is no supersymmetric vacuum, since $F_{\Phi_1} \neq 0$
always.

Suppose at a point in the moduli space, centered say at $\Phi_1 =0$,
the K\"ahler potential takes the form
$$
K = |\Phi_1|^2 + c |\Phi_1|^4 + \ldots
$$
where the higher order terms are suppressed by a
characteristic mass scale (which we set to one).
Then, provided:
$$
\qquad |c a^2| \gg 1,  \qquad c<0,
$$
the theory has a non-supersymmetric vacuum at
\eqn\nsv{
\Phi_{1,*} = {1\over c a^*},
}
which
breaks SUSY at an exponentially low scale.

This can be seen as follows. Expanded about small $\Phi_1$, the superpotential ${\tilde W}_1$ takes the form
$$
\tilde \CW_{1}(\Phi_1)=-{S\over2}\log(a-\Phi_1)/\Delta + \ldots
$$
where the subleading terms are suppressed by additional powers of $S$, but are otherwise regular at the origin of $\Phi_1$ space.
Integrating out $S$ first, by solving its $F$ term constraint, we find
$$
{S}_*=\Delta^3 exp(-{\tilde t}/g_s)+\ldots
$$
where
$$
\tilde{t} = t - {\half} g_s\log(a/\Delta)
$$
and the subleading terms are of order $\Phi_1/a$ which will turn out
to be small in the vacuum. For large $\tilde t$, $S$ is generically
very massive, so integrating it out is justified.

The potential for $\Phi_1$ now becomes
$$
V_{eff}(\Phi_1) = {1\over 1 + c|\Phi_1|^2}\,{|S_{*}|^2 \over |a-\Phi_1|^2}+\ldots
$$
It is easy to see that, up to corrections of order $1/|a^2 c|$ and
$S_*/(ma^2)$, this has a non-supersymmetric vacuum
at \nsv\
where $\Phi_1$ has a mass squared of order
$$
-c | {S_{*}\over a}|^2.
$$
This is positive, and the vacuum is (meta)stable, as long as $c<0$.
Note that we could have obtained the Polonyi model as a limit of the Fayet model where we turn on a very large FI term for the off-diagonal gauge group of nodes one and two. In this case, the stability of the Fayet model for a generic (effectively canonical) K\"ahler
potential guarantees that the Polonyi model obtained from it is stable.
In fact \AKS , as we'll review in section 7 , one can show this
directly
by computing the relevant correction to the K\"ahler potential,
arising from loops of massive gauge bosons.

\newsec{An O'Raifeartaigh model}

To represent the third simple classic class of SUSY breaking models,
we engineer
an O'Raifeartaigh model.
Consider the $A_3$ fibration with
\eqn\zsare{z_1(x) = mx,~z_2(x) = mx, ~z_3(x) = mx, ~z_4(x) = -m(x-2a)~.}
The defining equation of the non-compact Calabi-Yau
is then
\eqn\defe{uv = (z-mx) (z-mx) (z-mx) (z+m(x-2a))~.}

We wrap 1 D5 brane on each of $S^{2}_{1,2,3}$.  The adjoints $\Phi_1$
and $\Phi_2$ are massless, while $\Phi_3$ obtains a mass from its
superpotential
\eqn\wthree{\CW_3(x) = \int (z_3(x) - z_4(x)) ~dx }
which gives
$$
\CW_3(\Phi_3) = m (\Phi_3 - a)^2~.
$$
Of course, there are also quarks $Q_{12}, Q_{21}$ and $Q_{23}, Q_{32}$.
They couple via the superpotential couplings
\eqn\quarksup{Q_{12} \Phi_1 Q_{21} - Q_{12} \Phi_2 Q_{21} + Q_{23} \Phi_2
Q_{32} - Q_{23} \Phi_3 Q_{32}~.}
Because $\Phi_3$ is locked at $a$, for generic values of $\Phi_2$,
$Q_{23}$ and $Q_{32}$ are massive.  Then node 3 is entirely massive,
and we can perform a geometric transition.

The resulting theory has a new ``glueball superfield" $S$, and effective
superpotential
\eqn\neww{{\cal W}_{eff} = Q_{12} \Phi_1 Q_{21} - Q_{12} \Phi_2 Q_{21}
-{1\over 2} S {\rm \log}(a-\Phi_2)/\Delta
+ S ({\rm log} (S/\Delta^3) - 1) + {t\over g_s} S+\ldots}

Integrating out the $S$ field
yields (at leading order)
\eqn\sis{S_* = \Delta^3 e^{-{\tilde t}/g_s}~.}
where
$${\tilde t} = t + {1\over 2} g_s \log(a/\Delta).
$$
Plugging this into the superpotential yields:
\eqn\newtwo{{\cal W}_{eff} = Q_{12} \Phi_1 Q_{21} - Q_{12} \Phi_2 Q_{21}
- {1\over 2 } S_{*} \Phi_2/a~+\ldots}
The omitted terms are suppressed by more powers of $\Phi_2/a$.
We recognize
\newtwo\ as the superpotential for an O'Raifeartaigh model,
very similar to the one considered in \AKS.
We see
that setting $F_{\Phi_1}=F_{\Phi_2}=0$ is impossible, so one obtains
F-term supersymmetry breaking, with a small scale set by
$\Delta e^{-t/3 g_s}$.

The stability of the non-supersymmetric vacuum again depends on
the form of (technically) irrelevant corrections to the K\"ahler potential. As in the case of Polonyi model,
corrections which yield a stable vacuum can
be arranged by embedding the model in a slightly
larger theory. We'll turn to this in the next section.

\newsec{A Master Geometry}

It is possible to construct one configuration of branes on an $A_4$
geometry which in appropriate limits can be made to reduce to any of
the three simple models discussed in the previous sections.  The
geometry is described by the defining equation
\eqn\master{
uv=(z-mx)(z-mx)(z+mx)(z-mx)(z+m(x-2a))
}
which has superpotential given by 
\eqn\masterw{
\CW_{master}=\sum_{i=1}^{4}\CW_i(\Phi_i)+\sum_{i=1}^{3}{\rm
Tr}(Q_{i,i+1}\Phi_{i+1} Q_{i+1,i}- Q_{i+1,i}\Phi_iQ_{i,i+1}). }
where we wrap $N$ branes on nodes one, two and three, and a single
brane on node four.    The superpotentials
for the adjoints are given by
$$ \CW_1(\Phi_1)=0, ~~ \CW_2(\Phi_2)=-m {\rm Tr}
(\Phi_2^2), ~~ \CW_3(\Phi_3)=m {\rm Tr}(\Phi_3^2),~~
\CW_4(\Phi_4)=-m(\Phi_4-a)^2~.  $$
For simplicity of the discussion, we'll set $N=1$
in this section. The non-abelian generalization is immediate, since
all the nodes are asymptotically free (for large adjoint masses). 
As long as the scale of supersymmetry breaking driven by the geometric
transition is high enough, we can ignore the non-abelian gauge dynamics
on the other nodes.
\bigskip
\centerline{\epsfxsize 3.5truein\epsfbox{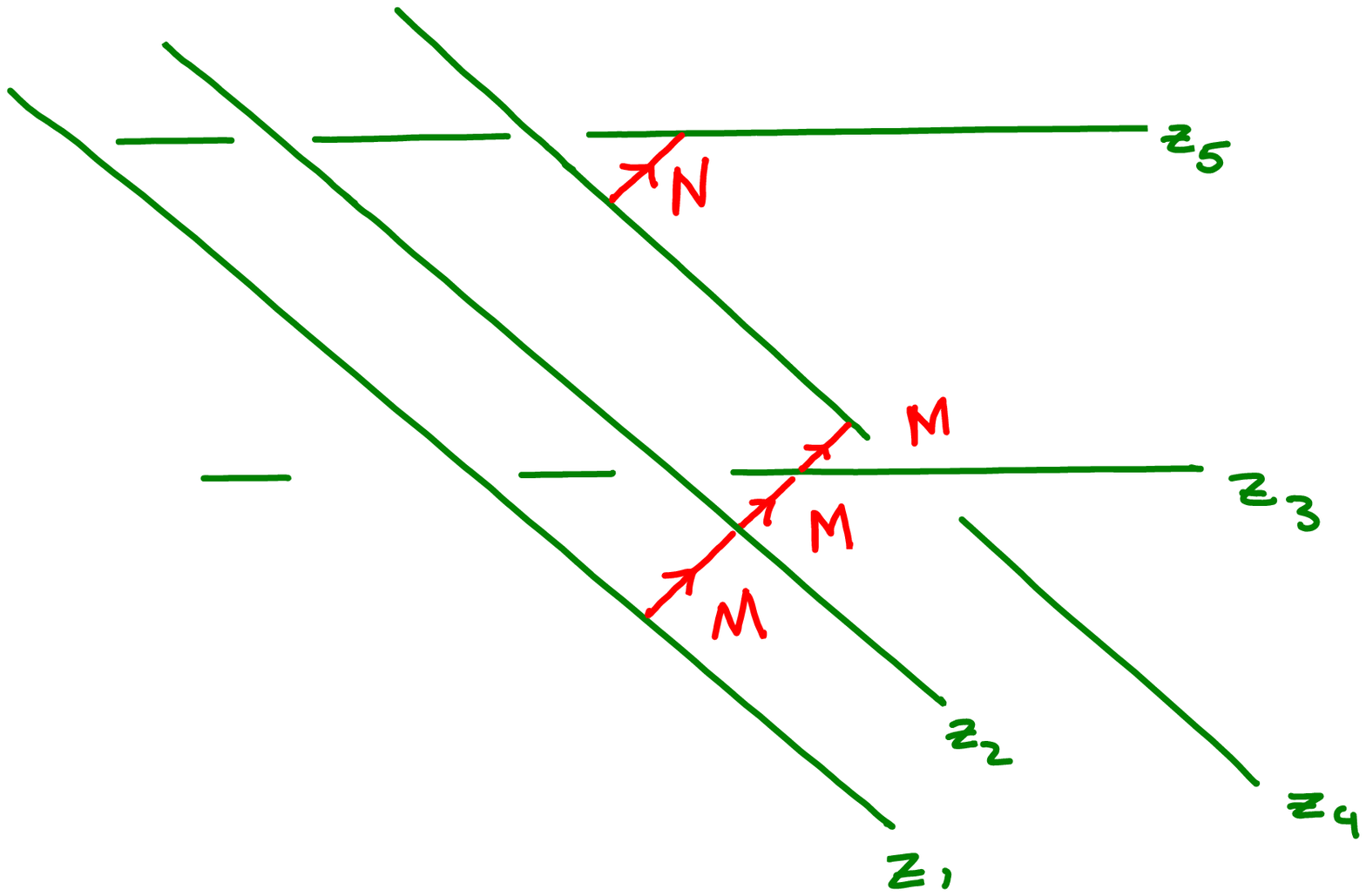}}
\noindent{\ninepoint
\baselineskip=2pt {\bf Fig. 2.} {
The master $A_4$ geometry that gives rise to Fayet, Polonyi and
O'Raifeartaigh models by turning on suitable FI terms.
The stringy instantons associated with the massive fourth node
generate the non-perturbative superpotential that triggers
dynamical supersymmetry breaking in the rest of the theory.
}}
\bigskip
The master theory has a metastable non-supersymmetric vacuum for
generic, non-zero FI terms. We can recover all three of the models discussed
above by introducing large Fayet-Iliopoulos terms for certain pairs of quarks,
so we expect that these will have non-supersymmetric vacua as well. This approach
to obtaining the canonical models is particularly useful in the case of Polonyi and O'Raifeartaigh models, for which we needed to assume a particular sign for the subleading correction to the K\"ahler potential. By obtaining the theories
from the master theory, we can compute the corrections to the K\"ahler
potential directly and show that they are of the type required to
stabilize the susy-breaking vacua.

To see that the master theory has a metastable non-supersymmetric
vacuum, we can proceed as in the Fayet model. Node four is massive,
and the corresponding non-perturbative superpotential
can be computed in the geometry after transition. The effective
superpotential after the transition and integrating out the massive
adjoints $\Phi_{2,3}$ is then easily seen to be
$$
\CW_{eff} = Q_{12}Q_{21}\Phi_1 + {S_{*}\over 4 ma}
(Q_{23}Q_{32} +\ldots)
$$
where we have omitted quartic and higher order terms in the $Q$'s which do
not affect the status of the vacuum. With generic FI terms setting 
$$
|Q_{12}|^2-|Q_{21}|^2 = r_2, \qquad  |Q_{23}|^2-|Q_{32}|^2 = r_3~,  
$$
this is easily seen to have an isolated vacuum which breaks
supersymmetry.

We'll now show that we can recover all of the the three models studied so far in particular regimes of large FI terms.

\subsec{O'Raifeartaigh}

We can recover the O'Raifeartaigh construction by turning on a large
FI term for $Q_{23}$ and $Q_{32}$ -- that is, for the U(1) under
which these are the only charged quarks.  This generates a D-term
\eqn\ODterm{ D_{O'R}= |Q_{23}|^2 -|Q_{32}|^2-r_3. }
Taking $r_3
>> 0$, this requires that $Q_{23}$
acquire a large expectation value.  Additionally there is an F-term
for $Q_{32}$ \eqn\OFterm{ F_{Q_{32}}=Q_{23}(\Phi_3-\Phi_2) } which,
in light of the D-term constraint, will set $\Phi_2$ equal to
$\Phi_3$.
The superpotential then becomes just the O'Raifeartaigh
superpotential of the previous section (with certain indices
renamed),
\eqn\OSuper{ \CW_{O'R}=m(\Phi_4-a)^2+Q_{12}
Q_{21}(\Phi_2-\Phi_1)+Q_{24} Q_{42}(\Phi_4-\Phi_2)~. }
By performing
a geometric transition on the massive node, we recover the superpotential
\neww.

\subsec{Fayet}

Alternatively, we could have turned on a large FI term for $Q_{12}$ and $Q_{21}$, generating a D-term
\eqn\FDterm{ D_{Fayet}=|Q_{12}|^2-|Q_{21}|^2-r_2. }
In conjunction with the F-term for $Q_{21}$, by the same process as in the O'Raifeartaigh model, $\Phi_1$ is set equal to $\Phi_2$.  This time, the remaining superpotential is given by
\eqn\FSuper{
\CW_{Fayet}=m\Phi_2^2-m\Phi_3^2+m(\Phi_4-a)^2+Q_{23} Q_{32}(\Phi_3-\Phi_2)
+\ldots
}
which is precisely the superpotential associated with the Fayet geometry \athree.  Performing a geometric transition on $S^2_4$, we recover the Fayet model as discussed in section 4.

\subsec{Polonyi}

From the Fayet model above, before the geometric transition, we can turn turn on another D-term for the quarks $Q_{23}$ and $Q_{32}$,
which along with the F-term for $Q_{32}$ sets $\Phi_2=\Phi_3$.  The superpotential becomes
$$
\CW=-m\Phi_3^2 + m(\Phi_4-a)^2+Q_{34}Q_{43}(\Phi_4-\Phi_3)
$$
which reproduces the Polonyi model of section 5.  Again performing the geometric transition on $S^2_4$ results in the actual Polonyi model.

\subsec{The K\"ahler potential}

The O'Raifeartaigh and Polonyi models have flat directions at tree
level.  As we discussed for e.g. the Polonyi model, the existence of a
stable SUSY-breaking vacuum depends on the sign of the leading,
quartic correction to the K\"ahler potential.  When we obtain the
model as a suitable limit of our master model as above, we can compute
this correction and verify explicitly that the vacuum is stable.  Let
us go through this in some detail. In fact, for simplicity, let's focus on obtaining a stable Polonyi
model as a limit of a Fayet model \AKS.

After the geometric transition in the Fayet model, the effective theory is characterized by a superpotential
\eqn\supnow{{\cal W} =
{S_*\over ma} Q_{23} Q_{32} +\ldots
}
and D-term
\eqn\Dis{D = |Q_{32}|^2 - |Q_{23}|^2 - r_3 ~.}
Here $r_3$ is the FI term for the $U(1)$ under which only $Q_{23}$ and $Q_{32}$ carry a charge. We can expand this theory about the vev $Q_{23} = \sqrt{r_3}$.  Renaming
$$X = Q_{32}~,
$$
the effective theory then has
\eqn\pollim{{\cal W} = {S_{*}\over ma} \sqrt{r_3} X~.}

To find the K\"ahler potential for $X$, we should integrate out the
massive $U(1)$ gauge multiplet.  What happens to the potential
contribution from the D-term, \Dis?  As explained in \Nima, in the
theory with the $U(1)$ gauge field, gauge invariance relates D-term
and F-term vevs at any critical point of the scalar potential.  When
one integrates out the $U(1)$ gauge field, there is a universal
quartic correction to the K\"ahler potential which (using the
relation) precisely reproduces the potential contribution from the
D-term.  For the theory in question, the quartic correction to the
K\"ahler potential for $X$ is just
\eqn\kahcor{\Delta K = -{g_{U(1)}^2 \over M_{U(1)}^2} (X^\dagger X)^2~.}
Here $M_{U(1)}$ is the mass of the $U(1)$ gauge boson, $M_{U(1)}
\sim g_{U(1)} \sqrt{r_3}$.
The result is a quartic correction to $K$
\eqn\result{\Delta K = - {1\over r_3} (X^\dagger X)^2~.}
So in the notation of section 5,
$$c = -{1\over r_3}$$
and the sign $c<0$ results in a stable vacuum, as expected.
Plugging in the $F$-term $F_X \sim {S_*\over ma}\sqrt{r_3}$,
\result\ gives $X$ a mass
$$m_{X} \sim {S_{*} \over ma},$$
in agreement with what it was in the full, Fayet model.
Note that while one would obtain other quartic couplings in $K$ after integrating out the $U(1)$ gauge boson, they don't play
any role.  They involve powers of the heavy field $Q_{34}$, and since
$F_{Q_{34}} \ll F_X,$
cross-couplings of the
form $Q_{34}^\dagger Q_{34} X^\dagger X$ in $K$ do not correct the estimate for the $X$ mass above appreciably.

\newsec{Generalization}

We now present a very general argument for the existence of supersymmetry-breaking effects in a class of stringy quiver gauge theories which includes those just discussed.  Suppose we have such an $A_r$ quiver theory in which the last node is isolated and undergoes a transition.  Note that this is the case in the master geometry considered in the previous section.

In this case, the transition deforms the geometry to the following:
$$
uv=\left(\prod_{i=1}^{r-1}(z-z_i(x))\right)\left((z-z_r(x))(z-z_{r+1}(x))-s\right)
$$
where 
in which case the superpotential for the branes on the second-to-last node becomes
$$
\tilde\CW_{r-1}(\Phi_{r-1})=\int dx(\tilde z_{r}(x)-z_{r-1}(x))
$$
where $\tilde z_{r}(x)$ is the solution to the equation
\eqn\deformed{
(z-z_r(x))(z-z_{r+1}(x))=s
}
which asymptotically approaches $z_r(x)$.  We can re-write the superpotential as a correction to the pre-transition superpotential as
$$
\tilde\CW_{r-1}(\Phi_{r-1})=\int dx (\tilde z_r(x)-z_r(x))+\CW_{r-1}(\Phi_{r-1})
$$ 
and the F-term for $\Phi_{r-1}$ and the remaining adjoints are then given by
\eqn\fgen{\eqalign{
F_{\Phi_{r-1}}&=\CW_{r-1}^\prime(\Phi_{r-1})+(\tilde z_r(\Phi)-z_r(\Phi))+Q_{r-1,r} Q_{r,r-1}\cr
F_{\Phi_i}&=\CW_{i}^\prime(\Phi_i)+Q_{i-1,i} Q_{i,i-1}-Q_{i,i+1} Q_{i+1,i}\cr
}}
which we can combine to obtain the constraint
\eqn\const{
\sum_i^{r-1}\CW_{i}^\prime(\Phi_i)=z_r(\Phi_{r-1})-\tilde z_r(\Phi_{r-1})
}
Note that the right hand side here cannot vanish for any value of $\Phi_{r-1}$ since $z_r(x)$ 
can never solve \deformed, the solution to which defines $\tilde z_r(x)$

If we now consider turning on generic FI terms for the $U(1)$ gauge groups, the D-term constraints will require that, say, the $Q_{i,i+1}$'s acquire vevs while the $Q_{i+1,i}$'s get fixed at zero.  The F-terms for the $Q_{i+1,i}$'s will then in turn require
$$
\Phi_i=\Phi_j
$$
for all $i,j$.  When the brane superpotentials for the first $r-1$ nodes 
are of the form
$$
\CW_{i}(\Phi_i) = \epsilon_i m \,\Phi_i^2, \qquad i=1,\ldots r-1.
$$
where $\epsilon_{i} =0 \pm 1$, the left hand side of \const\ vanishes, 
while the right hand side is
strictly non-zero. It is exponentially small, as long as the last node was
isolated 
$$
\CW_{r}(\Phi_r) = m(\Phi_r-a)^2
$$
before the transition.  This generically triggers low-scale susy breaking.

In terms of the classic models discussed in this paper, one can immediately see that the susy breaking in the Fayet model and in the master geometry can be explained by the above analysis.  In the case of the Polonyi and O'Raifeartaigh models, it is even simpler, since the left hand side of \const\ vanishes {\it identically} for those models.
One could conduct a similar analysis for configurations with more complicated superpotentials and non-generic F-terms on a case-by-case basis.  What we see is that often the susy-breaking effects caused by the geometric 
transition can be understood at an exact level.

\newsec{SUSY breaking by the rank condition}

Here, we exhibit models which break supersymmetry due to the ``rank condition."  This
class of models is very similar to those arising in studies of metastable vacua in SUSY
QCD \ISS.  However, we work directly with the analogue of the magnetic dual variables,
and the small scale of SUSY breaking is guaranteed by retrofitting \DFS.

Consider the $A_3$ fibration with
\eqn\choice{z_1(x) = mx, \qquad z_2(x) = -mx, \qquad z_3(x) = -mx, \qquad z_4(x) = - m(x-2a)~.}
Then the defining equation is
\eqn\defis{uv = (z-mx) (z+mx) (z+mx) (z + m (x-2a))~.}
We choose to wrap $N_f - N_c$ D5 branes on $S^2_1$, $N_f$ D5 branes on $S^2_2$, and a single D5
on $S^2_3$.
The tree level superpotential is
\eqn\tresup{
\CW=\sum_{i=1}^{3}\CW_i(\Phi_i)+\sum_{i=1}^{2}(Q_{i,i+1}\Phi_{i+1} Q_{i+1,i}- Q_{i+1,i}\Phi_iQ_{i,i+1}).
}
where
$$
\CW_1(\Phi_1)=m {\rm Tr}(\Phi_1)^2,\qquad
\CW_2(\Phi_2)=0, \qquad
\CW_3(\Phi_3)=- m (\Phi_3-a)^2
$$
\bigskip
\centerline{\epsfxsize 2.5truein\epsfbox{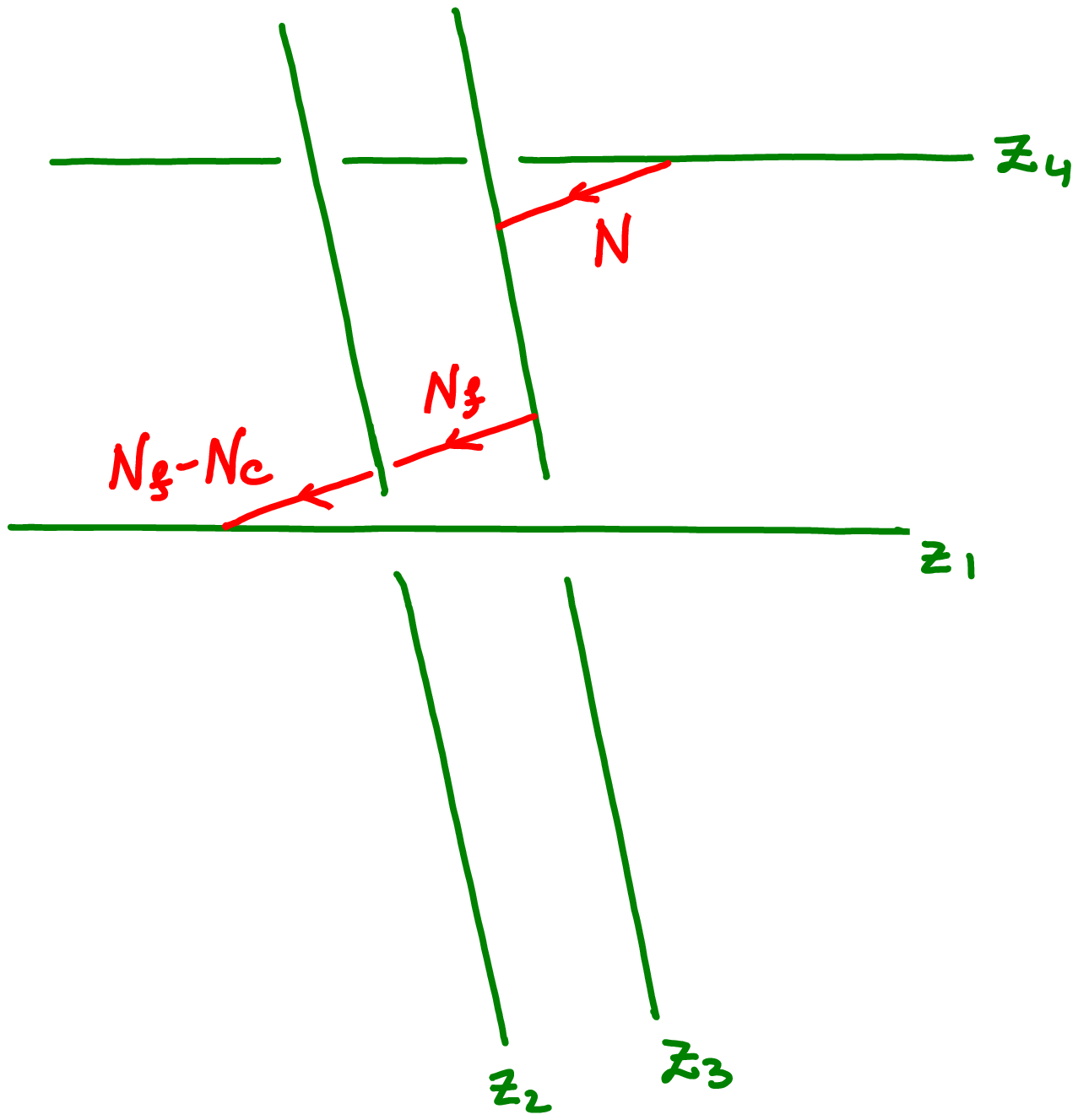}}
\noindent{\ninepoint
\baselineskip=2pt {\bf Fig. 3.} {
The (magnetic) $A_3$ geometry that retrofits the ISS model.
}}
\bigskip
Now, we replace the third ($U(1)$) node with an $S^3$ with flux, and
integrate out $\Phi_1$ trivially (we can take the mass to be very
large).  The result is:
\eqn\rank{ \CW = S (\log{(S/Delta^3)} - 1) + {t\over g_s}S  - {1\over 2} S~ {\rm Tr} {\rm log}(a-\Phi_2)/\Delta
- Q_{12} \Phi_2 Q_{21}~+\ldots }
where the omitted terms are suppressed by additional powers of $S$.
Integrating out $S$ in a Taylor expansion about $\Phi_2=0$, produces a 
theory with superpotential
\eqn\rankbre{\CW = S_{*} {\rm Tr}\Phi_2/a - {\rm Tr} Q_{12} \Phi_2 Q_{21}~+\ldots}
where 
\eqn\SS{S_{*} = \Delta^3 exp(-{\tilde t/g_s})~,}
and $\tilde t = t - N_f{1\over 2} g_s \log(a/\Delta)$.
Computing $F_{\Phi_2}$, we see that the contribution from the first term in \rankbre\ has rank $N_f$,
while the contribution from the second term has maximal rank $N_f - N_c < N_f$.  The two cannot cancel
and SUSY is broken.  However, due to the small coefficient of the ${\rm Tr}\Phi_2$ term, the breaking occurs
at an exponentially small scale.

This model is very similar to the theories analyzed in \ISS\ (for
$N_{c}+1 \leq N_f < {3\over 2} N_c$) and in section 4 of \Giveon.  One
difference is that the origin of the small parameter is dynamically
explained.  The discussion of corrections due to gauging of the
$U(N_f)$ factor (which is a global group in \ISS) is identical to that
in \Giveon\ up to a change of notation, and we will not repeat it
here.  For large $a$, the higher order corrections to \rankbre\ (which
are suppressed by powers of $\Phi_2/a$) should not destabilize the
vacuum at the origin, described in \refs{\ISS,\Giveon}.

We could also replace the $U(1)$ at node 3 with a $U(N)$ gauge group,
still in the same geometry.  Then, in \rank, the coefficient of the $S
{\rm log} S$ term is changed to $N$.  The only effect, after a geometric
transition at node 3, is the replacement of
replacement $e^{-t/g_s} \to e^{-t/g_s N}$ in \SS .
This model, where the node upon which we perform the geometric
transition has non-Abelian gauge dynamics, is a literal example of the
retrofitting constructions of \DFS.  The $\Phi_2$ field appears in the
gauge coupling function of the $U(N)$ gauge group at node 3, because
it controls the masses of the quarks $Q_{23}$ and $Q_{32}$ which are
charged under $U(N)$.  At energies below the quark mass, the $U(N)$
is a pure ${\cal N}=1$ gauge theory and produces a gaugino
condensation contribution $\Lambda_{N}^3$ in the superpotential.  The
standard result for matching the dynamical scale of the low-energy
pure $U(N)$ theory to the scale $\Lambda_{N,N_f}$ of the higher
energy theory with $N_f$ quark flavors with mass matrix $\tilde m$
is\foot{Here, we are assuming the adjoints are very massive $m \to
\infty$ and are just matching the QCD theories with quark flavors.}
\eqn\matching{\Lambda_{N}^{3N} = \Lambda_{N,N_f}^{3N-N_f} {\rm det} \tilde m~.}
With the identification of $S$ with the gaugino condensate \Vafa\
$$S \sim tr(W_{\alpha}^2) = \Lambda_{N}^3,
$$
and identifying
the mass matrix $\tilde m = a - \Phi_2$, we see that we predict
\eqn\sis{S^N = \Lambda_{N,N_f}^{3N-N_f} {\rm det} (a-\Phi_2)~.}
This is precisely what carefully integrating $S$ out of \rank\ produces,
with $\Lambda_{N,N_f}^{3N-N_f} = \Delta^{3N-N_f} e^{-t/g_s}$.
So in our model with $N>1$, the small ${\rm Tr}(\Phi_2)$ term in
\rankbre\ can really be thought of as arising from the presence of $\Phi_2$ in the gauge
coupling function for the $U(N)$ factor.

\newsec{Orientifold models}

In the presence of orientifold 5-planes, we expect D1 brane instantons
wrapping 2-cycles that map to themselves to contribute to the
superpotential. The D1 brane instanton contributions should again be
computable using a geometric transition that shrinks
the $S^2$, and replaces it with an
$S^3$. Geometric transitions with orientifolds have been studied for e.g. in \refs{\sinha,\IKV}.

After the transition we generally get 2 different contributions to the
superpotential. First, the charge conservation of the D5/O5 brane
that disappear after the transition, requires a flux through the
$S^3$ equal to the the amount of brane charge:
$$
\CW_{flux} = {t\over g_s}\, S + N_{D5/O5} \del_{S} {\cal F}_0
$$
Second, there can be additional O5 planes that survive as the
fixed points of the holomorphic involution after the transition.
The O5 planes, just like D5 branes generate a superpotential \AAHV\
$$
\CW_{O5} = \int_{\Sigma} \Omega,
$$
where the integral is over a
three-chain with a boundary on the orientifold plane.  The
contributions to the superpotential due to O5 planes and RR flux of
the orientifold planes are both computed by topological string $RP^2$
diagrams. The contributions of physical brane charge come from the
sphere diagrams.

In this way, geometric transitions can be used to sum up the instanton
generated superpotentials in orientifold models. Analogously to our
discussion of the previous sections, this can be used for dynamical
supersymmetry breaking. We'll discuss in detail the Fayet model below;
others can be seen to follow in similar ways.

\subsec{The Fayet model}

Consider orientifolding the theory from section 3, by combining the
worldsheet orientation reversal with an involution $I$ of the
Calabi-Yau manifold. For this to preserve the same supersymmetry as
the D5 branes, the holomorphic involution $I$
of the Calabi-Yau has to preserve the holomorphic three-form $\Omega=
du/u dzdx = - dv/v dz dx$.

An example of such an involution is one that takes
$$
x \rightarrow -x
$$
and
$$
u \rightarrow v, \qquad v \rightarrow u
$$

A simplest Fayet-type model built on this orientifold is an $A_5$
geometry that is
roughly a doubling of that in section 4:
$$
uv=(z-mx)^2(z+mx)^2(z-m(x-2a))(z+m(x-2a)).
$$
We'll blow this up in a sequence:
$${\eqalign{
&z_1(x)=mx, \qquad z_2(x) = -m(x-2a), \qquad z_3(x) =mx, \;\;\cr
&z_4(x)=-mx,\qquad
z_5(x)= +m(x+2a),\qquad z_6(x)=-mx, }}
$$
It can be shown that the orientifold projection ends up mapping
$$
S^2_i \rightarrow S^2_{6-i},
$$
fixing $S^2_3$. Consider wrapping $M$ branes on $S^2_{i}$ for
$i=1,2$, and their mirror images, and $2N$ branes on $S^2_3$. With a particular choice
of orientifold projection, the gauge group on the branes is going to be
$$
U(M)\times U(M) \times Sp(N)$$
Since the orientifold flips the sign of $x$, on the fixed node $S^2_3$
it converts $\Phi_3$ to an adjoint of $Sp(N)$.
(Having chosen that the
orientifold sends $x$ to minus itself, the action on the rest of the
variables is fixed by asking that it preserve the same susy as the D5
branes, $and$ that it be a symmetry after blowing up.)
In the model at hand, the tree-level superpotential is
$$
\CW= \sum_{i=1}^3\CW_i(\Phi_i) + {\rm Tr}(Q_{12}\Phi_2Q_{21}
-Q_{21}\Phi_1 Q_{12}) +{\rm Tr}(Q_{23}\Phi_3
Q_{32}-Q_{32}\Phi_2 Q_{23}).
$$
where
$$
\CW_1(\Phi_1) = m {\rm Tr}(\Phi_1-a)^2, \qquad \CW_2(\Phi_2)=-m {\rm
Tr}(\Phi_2-a)^2,\qquad \CW_3(\Phi_3)=m {\rm Tr}\Phi_3^2.
$$
Note that, even though the $\IP^1$ is
fixed by the orientifold action, it is not fixed point-wise.
This means there is no O$5^+$ plane charge on it. Instead, there are two
$non-compact$ orientifold 5-planes.
This model is T-dual \lazaroiu\ to the O6-plane models of \Kachru .

After the geometric transition that shrinks node three and replaces it with an $S^3$
$$
S^2_3 \rightarrow S^3
$$
the geometry becomes:
$$
uv=(z-mx)(z+mx)(z-m(x-2a))^2(z+m(x-2a))^2
((z-mx)(z+mx)-s).
$$
where
$$
\int_{S^3}  \Omega
= S
$$
with $S=s/m$.
Since the orientation reversal acted freely on the $S^2_3$, there are only
$N$ units of D5 flux through the $S^3$
$$
\int_{S^3} H^{RR} = N
$$
which gives a superpotential
$$
\CW_{flux} = {t\over 2 g_s}\; S + N S(\log{S\over\Delta^3}-1)
$$
the overall factor of $1/2$ comes from the fact that
both the charge on the $S^2$ and its size has been cut in half by the orientifolding. Above, $t = \int_{S^2_3} k+i B^{RR}$ is the combination of 
Kahler moduli that survives the orientifold projection. 
In addition, the two non-compact O$5^+$
planes get pushed through the transition.
Because the space still needs 2 blowups to be smooth, to give a precise description of the O5 planes would require using a geometry covered with 4 patches.
At the end of the day, effectively, the O5 planes correspond to non-compact curves over the two points on the Riemann surface
$$
(z-{\tilde z}_{3}(x)) (z-{\tilde z}_{4}(x))=
((z-mx)(z+mx)-s)=0.
$$
located at $x=0$, and the corresponding values of $z$, $z_{\pm}(0)$.   They generate a superpotential
$$
\CW_{O5^+} = \int^{z_{-}(0)}({\tilde z}_{3}-{\tilde z}_4) dx+\int^{z_{+}(0)}
({\tilde z}_{3}-{\tilde z}_4) dx.
$$
One can show that the contribution of the O5 planes
is
$$
\CW_{O5^+} =
+S(\log{S\over\Delta^3}-1)
$$
The fact that the $RP^2$ contribution is proportional to that of the sphere is not an accident. It has been shown generally that the contribution of the O5 planes in these classes of models is
$ \pm \del_{S} {\cal F}_{S^2}$ \refs{\oz,\IKV}.
This means that the O5 planes and the fluxes add up to
$$
N+1
$$
units of an ``effective'' flux on the $S^3$.

After the transition, the branes on node three have disappeared and
with them $\Phi_3$ and $Q_{23}, Q_{32}$. In addition, the
deformation of the geometry induces a deformation of the
superpotential for node 2:
$$
{\tilde \CW}_2(x)=\int(z_2(x)-{\tilde z}_3(x)) dx
$$
where one picks for ${\tilde z}_3$ the root that asymptotes to $+mx$.
This is
$$
{\tilde \CW}_2(x)=\int (- m(x-2a) - \sqrt{(mx)^2+s}) dx,
$$
which, when expanded near the vacuum at $x=a$, gives
$$
{\tilde \CW}_2(\Phi_2)=-{\rm Tr}\,  m(\Phi_2-a)^2 - \half S \,{\rm Tr}
\log( \Phi_2/\Delta)+\ldots
$$

The effective superpotential that sums up the instantons is thus
$$
{\cal W}_{eff}=
\CW_1(\Phi_1)+{\tilde W}_2(\Phi_2,S)
+
{
\rm Tr}(Q_{12}\Phi_2 Q_{21}
-Q_{21}\Phi_1 Q_{12})
+\CW_{flux}+\CW_{O5}
$$
Up to an overall shift of both $\Phi_{1,2}$ by $a$, this is the
same model as in section 3.

We expect a transition here even when $N=0$, and there are no D5
branes on the $S^2$. The transition for $Sp(0)$ is analogous to the
transition that occurs for a single D-brane on the $S^2$, and a $U(1)$
gauge theory. In both cases, the smooth joining of the $S^2$ and the
$S^3$ phases is due to instantons that correct the geometry. In the
orientifold case at hand, it is important to note that, while there is
no flux through the $S^3$, the D3 brane wrapping it is absent: the
orientifold projection projects out \hori\ the ${\cal N}=1$ $U(1)$ vector
multiplet associated with the $S^3$, and with it the D3 brane charged
under it.

Picking the other orientifold projection, the $Sp(N)$ gauge group
gets replaced with an $SO(2N)$ with $\Phi_3$ becoming the
corresponding adjoint. In this case, much of the story remains the
same, except that the $RP^2$ contribution becomes
$$
\CW_{O5^-} =
-S(\log {S\over\Delta^3}-1).
$$
This means that the O$5^-$ planes and the fluxes add up to
$$
N-1
$$
units of an ``effective'' flux on the $S^3$.
This is negative or zero for $N\leq 1$. Naively, the negative effective flux
breaks supersymmetry after the transition. This is clearly impossible.
It has been argued in \IKV\ that the correct interpretation of this is that in fact
$SO(2)$, $SO(1)$ and $SO(0)$ cases do not
undergo the geometric transition. This has to correspond to the statement that in these cases there are no D1 brane instantons on node three,
and that the classical picture is $exact$ in these cases. This translates in the statement that in these cases, in
$$
{\cal W}_{eff} = {\cal W}_{eff}|_{S=0}
$$
$S$
should not be extremised, but rather set to zero identically
in the effective superpotential.

Note that with the $SO$ projection on the space-filling branes, a D-instanton wrapping the same node
enjoys an $Sp$ projection.  As discussed in \refs{\zeromodes,\Kachru}, in this situation direct zero-mode
counting also suggests that the instanton should $not$ correct the superpotential.
There are more than two fermion zero modes coming from the Ramond sector of strings stretching from the instanton
to itself.  This is in accord with the results of \IKV.
In contrast, when one has an $Sp$ projection on the space-filling branes, the instanton
receives an $SO$ projection, and the instanton with $SO(1)$ worldvolume gauge group has the correct
zero mode count to contribute.  The presence of the instanton effects when one has this projection
(and their absence when one does not), was also confirmed by direct studies of the renormalization
group cascade ending in the appropriate geometry in \Kachru.

\medskip
\centerline{\bf{Acknowledgements}}
\medskip
We would like to thank B. Florea, B. Freivogel, T. Grimm, J. McGreevy, D. Poland, N. Saulina, E. Silverstein and C. Vafa for helpful discussions.
The research of M.A. and C.B. was supported in part by the UC
Berkeley Center for Theoretical Physics.  The research of M.A.
is also supported by a DOE OJI Award, the Alfred P. Sloan Fellowship, and
NSF grant PHY-0457317.
S.K. was supported in part by the Stanford Institute for
Theoretical Physics, and by NSF grant PHY-0244728
and DOE contract DE-AC003-76SF00515.
S.K. acknowledges the kind hospitality of the MIT Center for
Theoretical Physics during the completion of this work.

\appendix{A}{Brane superpotentials}

We can compute the superpotential $\CW(\Phi)$
as function of the wrapped 2-cycles
$\Sigma$ by using the superpotential
\refs{\Witten,\Mina}
$$
\CW=\int_\CC\Omega
$$
where $C$ is a three-chain with one boundary being $\Sigma$ and the
other being a reference two-cycle $\Sigma_0$ in the same homology
class. It is easy to show \Witten\ that the critical points of the superpotential are holomorphic curves.  We will evaluate it for the geometries at hand.
We can write the holomorphic three-form of the non-compact Calabi-Yau in the usual way,
\eqn\holo{
\Omega={dv\wedge dz\wedge dx\over{dF\over du}}={dv\over v}\wedge dz \wedge dx.
}
Now for fixed values of $x$ and $z$, the equation for the CY threefold becomes $uv=const$, which is the equation for a cylinder.  By shifting the definition of $u$ or $v$ by a phase, we can insist that the constant is purely real, and then by writing $u=x+iy$, $v=x-iy$, the equation can be reformulated as two real equations in terms of the real $(x_R,y_R)$ and imaginary $(x_I,y_I)$ parts of $x$ and $y$.
\eqn\realeq{
x_R^2+y_R^2=C+x_I^2+y_I^2,\qquad x_Rx_I=y_Ry_I.
}
The first of these can be solved for any given values of $x_I$ and $y_I$ to give an $S^1$.  The second equation restricts the possible values which we choose for $x_I$ and $y_I$ to a one-dimensional curve in the $(x_I,y_I)$ plane, and so we have the topology of $S^1\times {\rm R}$, where the size of the $S^1$ degenerates at the points where $z=z_i(x)$ for any $i$.  By simultaneously shifting the phases of $u$ and $v$ according to
$$\eqalign{
u&\rightarrow e^{i\theta}u\cr
v&\rightarrow e^{-i\theta}v
}$$
the equation for the cylinder remains unchanged, and we simply rotate about the $S^1$ factor.  We can thus integrate $\Omega$ around the circle and obtain
$$
\int_{S^1}\Omega=dz\wedge dx
$$
up to an overall constant.  Now the $\IP^1$'s on which we are wrapping the D5 branes are the product of the $S^1$ just discussed and an interval in the $z$ direction between values where the $S^1$ fiber degenerates.  Thus, for a given $\IP^1$ class in which the vanishing $S^1$ occurs for $z_i(x)$ and $z_j(x)$, we can integrate $dz\wedge dx$ over the interval in the $z$-plane and obtain
$$
\int_{S^1\times I_{ij}}\Omega=(z_i(x)-z_j(x))dx.
$$
The superpotential for the D-branes then becomes a superpotential for the location of the branes on the $t$-plane.  Defining an arbitrary reference point $t_*$, we then have
\eqn\super{
\CW(x)=\int_{t_*}^t(z_i(x)-z_j(x))dx
}
Of course, the contribution to the superpotential coming from the limit of integration at $t_*$ is just an arbitrary constant and is not physically relevant.  Thus we write \super\ instead as the indefinite integral
\eqn\superp{
\CW(x)=\int (z_i(x)-z_j(x))dx.
}
\appendix{B}{Multi-instanton contributions}

In this appendix we demonstrate the computation of multi-instanton
corrections to the superpotential using the Polonyi model of section
5 as an example.  All the information about these corrections is
contained in the deformed superpotential for $\Phi$, \eqn\appdef{
\tilde \CW(x)=\int\left(m(x-a)-\sqrt{m^2(x-a)^2+m S}\right)dx }
along with the flux superpotential\foot{In the case of the Polonyi
model these two terms constitute the entire superpotential.  In the
more general case, however, there will be more fields with
superpotential terms, but it will remain the case that only these
two contributions play a role in determining instanton corrections.}
\eqn\appflux{ \CW_{flux}={t\over
g_s}S+S\left(\log{S\over\Delta^3}-1\right). } where the scale
$\Delta$ is determined by the one-loop contributions to the matrix
model free energy. The models considered in this paper are
particularly convenient since the purely quadratic superpotential
for the massive adjoint at the transition node guarantees that the
flux superpotential will be exact at one-loop order in the
associated matrix model \DV.

Extremizing the flux superpotential and expanding in powers of the
instanton action
$$
S_{inst}\sim\exp(-t/N g_s),
$$
we can determine multi-instanton contributions to a given
superpotential term.  Summing up the series contributing to a given
$\Phi^k$ term in will correspond to computing corrections to a
fixed, explicit disc diagram, and so we might expect these series to
exhibit some integrality properties.

We first expand the deformed superpotential $\CW_1(\Phi)$ as a power
series in the glueball superfield $S$, \eqn\series{
\tilde\CW(\Phi)=\int \left(m(x-a)-m(x-a)(1+\sum_{n=1}^{\infty}
{(-1)^{n-1} n (2n-2)! \over 2^{2n-1} (n!)^2} y^n) \right)dx } where
the expansion parameter $y$ can also be expanded as a power series
in $x$, \eqn\appexp{ y = {S\over m(x-a)^2}={S\over m
a^2}\left(1+\sum_{n=1}^\infty(n+1)(-1)^n\left({x\over
a}\right)^n\right). } We can integrate \series\ term by term to
obtain an expansion of the effective superpotential in powers of
$\Phi$.  However, it will be useful to represent this schematically
$$
\CW_{1}(\Phi)=c_1\Tr\Phi+c_2\Tr\Phi^2+\ldots\qquad\qquad
c_i=\sum_{n=1}^\infty c_i^{(n)}S^n
$$
where the coefficients $c_i$ are themselves written as power series
in $S$. Extremizing the superpotential with respect to $S$ gives an
equation for the values of $S$ \eqn\appF{
\log{S\over\Delta^3}=-{t\over
g_s}-\sum_{n=1}^{\infty}\sum_{i=1}^\infty
n~c_i^{(n)}S^{n-1}\Tr\Phi^i } which can be solved perturbatively in
powers of $S_{inst}$.  Re-inserting the resulting values into the
original superpotential then allows us to read off the
instanton-corrected superpotential of the low energy theory up to
any given number of instantons.  Below we display the linear and
quadratic terms at the three-instanton level.
$$
\CW_{eff}=\mu\Tr\Phi+m\Tr\Phi^2
$$
where \eqn\final{\eqalign{ \mu&={1\over2}{\Delta^3\over
a}e^{-{t\over g_s}}-{1\over8}{\Delta^6\over ma^3}e^{-{2t\over
g_s}}+{1\over16}{\Delta^9\over m^2a^5}e^{-{3t\over g_s}}+\ldots\cr
m&=-{7\over8}{\Delta^3\over a^2}e^{-{t\over
g_s}}+{11\over16}{\Delta^6\over ma^4}e^{-{2t\over
g_s}}+{1\over32}{\Delta^9\over m^2a^6}e^{-{3t\over g_s}}+\ldots\cr
}}
\smallskip
\noindent It may be interesting to see if there is some way to relate these to the exact formulae for multicovers derived in the resolution of the singularity in hypermultiplet moduli space when a 2-cycle shrinks in IIB string theory, given (up to mirror symmetry) in \conifold.

\listrefs
\bye